\documentclass[]{pasj02} 
\usepackage{comment}
\usepackage{url}
\newcommand{\rev}[1]{\textcolor{black}{#1}}

\jyear{2026}
\Received{}
\Accepted{}

\begin{document} 

\title{Time-Resolved Connection between Starspots and Flares in Nearby Young Solar-type Stars Observed by TESS}

\author{
 Daijiro \textsc{Hitotsuyanagi},\altaffilmark{1}\altemailmark \email{hitotsuyanagi.daijiro.23f@st.kyoto-u.ac.jp}
 Hiroto \textsc{Yamada},\altaffilmark{1}
 Daisuke \textsc{Yamashiki},\altaffilmark{1}
 Kazuma \textsc{Ishihara},\altaffilmark{1}
 Yosuke \textsc{Yamashiki},\altaffilmark{2}
 and 
 Kosuke \textsc{Namekata}\altaffilmark{3,4,5,6}
 \orcid{0000-0002-1297-9485} 
}
\altaffiltext{1}{Department of Physics, Faculty of Science, Kyoto University, Kitashirakawa-Oiwake-cho, Sakyo-ku, Kyoto 606-8502, Japan}
\altaffiltext{2}{Graduate School of Advanced Integrated Studies in Human Survivability (GSAIS), Kyoto University, Yoshida-Nakaadachi-cho, Sakyo-ku, Kyoto 606-8306, Japan}
\altaffiltext{3}{Heliophysics Science Division, NASA Goddard Space Flight Center, 8800 Greenbelt Road, Greenbelt, MD 20771, USA}
\altaffiltext{4}{The Catholic University of America, 620 Michigan Avenue, N.E. Washington, DC 20064, USA}
\altaffiltext{5}{The Hakubi Center for Advanced Research, Kyoto University, Yoshida-Honmachi, Sakyo-ku, Kyoto 606-8501, Japan}
\altaffiltext{6}{Department of Physics, Kyoto University, Kitashirakawa-Oiwake-cho, Sakyo-ku, Kyoto, 606-8502, Japan}




\maketitle

\begin{abstract}
Superflares are energetic explosions on stellar surface with energies of $10^{33}$--$10^{36}$ erg, significantly exceeding those of typical solar flares. While previous studies have suggested that these events are driven by magnetic energy stored in large starspots, the detailed time-resolved relationship between starspot area and flare activity on individual stars has remained unclear. In this paper, we investigate the time evolution of magnetic activity on three representative young solar-type stars (EK Draconis, DS Tucanae A, and V889 Herculis) using $\sim$7 years of photometric data from the Transiting Exoplanet Survey Satellite (TESS). 
We automatically detected stellar flares and derived the flare frequency, starspot area, and rotational period for each TESS sector covering $\sim$27 days.
As a result, we found that the flare frequency and starspot area vary significantly across sectors, although we could not identify any activity-cycle-like pattern. 
There is a positive correlation between the starspot area and flare occurrence frequency for all three targets and the power-law dependence is consistent among the stars. 
This result supports the physical picture that superflares on young solar-type stars are powered by magnetic energy stored in large starspots, analogous to solar flares, and that the energy release rate changes as the total stored magnetic energy varies. 
Furthermore, from the analysis of EK Draconis, we find a possible dependence of starspot area on rotation period, which may suggest that large starspots preferentially form at mid-latitudes. These findings demonstrate that the magnetic activity mechanisms established for the Sun extend to the extreme magnetic activity observed on young active stars.

\end{abstract}


\section{Introduction}

\begin{table*}[!tp]
\caption{Stellar parameters.\label{tab:targets}}
\centering
\begin{tabular}{lccc}
\hline
Parameter & V889~Her (TIC~471000657) & DS~Tuc (TIC~410214986) & EK~Dra (TIC~159613900)\\
\hline
Spectral Type &
G0V$^{a}$ &
G6V$\pm1$/K3V$\pm1^{f}$ &
G1.5V$^{c}$ \\

$V_{\rm mag}$ &
$7.45\pm0.04^{a}$ &
$8.55\pm0.01/9.65\pm0.03^{f}$ &
$7.60\pm0.01^{d}$ \\

Age (Myr) &
$30^{a}$ &
$45\pm4^{g}$ &
50--125$^{c,h}$ \\

$T_{\rm eff}$ (K) &
$5830\pm50^{a}$ &
$5430\pm80/4700\pm90^{f}$ &
5560--5750$^{c,e,i}$ \\

Radius ($R_\odot$) &
$1.09\pm0.05^{a}$ &
$0.96\pm0.03/0.86\pm0.04^{f}$ &
$0.94\pm0.07^{c}$ \\

Mass ($M_\odot$) &
$1.06\pm0.02^{a}$ &
$1.01\pm0.06/0.84\pm0.06^{f}$ &
$0.95\pm0.04^{c}$ \\

Distance (pc) &
$35.36\pm0.02^{b}$ &
$44.78\pm0.03^{b}$ &
$34.40\pm0.03^{b}$ \\

$P_{\rm rot}$ (d) &
$1.3371\pm0.0002^{a}$ &
$2.85^{+0.04}_{-0.05}/-^{f}$ &
$2.766\pm0.002^{c}$ \\

$v\sin i$ (km\,s$^{-1}$) &
$39.0\pm0.5^{a}$ &
$17.8\pm0.2/14.4\pm0.3^{f}$ &
$16.4\pm0.1^{c}$ \\

Inclination (deg) &
$\approx55^{a}$ &
$>82^\circ/-^{f}$ &
$60\pm5^{c}$ \\

Binarity &
single$^{a}$ &
binary / exoplanet system &
low-mass companion$^{c}$ \\

\hline
\end{tabular}

\vspace{1mm}
\begin{minipage}{0.9\textwidth}
\footnotesize
$^{a}$ Table~2 of \citet{2003A&A...411..595S}. $^{b}$ Gaia EDR3 \citep{2021A&A...649A...1G}. $^{c}$ \citet{2017MNRAS.465.2076W}. $^{d}$ \citet{2000A&A...355L..27H}. $^{e}$ \citet{2018A&A...620A.162J}. $^{f}$ \citet{2019ApJ...880L..17N}. $^{g}$ \citet{2015MNRAS.454..593B}. $^{h}$ Reported ages for EK~Dra range from 30--125 Myr depending on the study. $^{i}$ For flare-energy calculations we adopt $T_{\rm eff}\approx5700$\,K from \citet{2005A&A...435..215K}.
\end{minipage}

\end{table*}

Solar flares are energetic explosions on the solar surface caused by the sudden release of magnetic energy stored around sunspots via magnetic reconnection \citep{2011LRSP....8....6S,2010ARA&A..48..241B,2024LRSP...21....1K}.
While typical solar flares have energies of $10^{29}$--$10^{32}$ erg \rev{(\cite{2012ApJ...759...71E,2022LRSP...19....2C})}, the Kepler mission \rev{\citep{2010Sci...327..977B}} has revealed that solar-type stars (G-type main-sequence stars) can produce ``superflares'' with energies ranging from $10^{33}$ to $10^{36}$ erg (e.g., \cite{2012Natur.485..478M,2013ApJS..209....5S,2016ApJ...829...23D,2021ApJ...906...72O,2022PASJ...74.1295Y,2024Sci...386.1301V}).
Subsequent studies have suggested that these superflares are also driven by magnetic energy stored in large starspots (e.g., \cite{2013PASJ...65...49S,2013ApJ...771..127N,2015PASJ...67...33N,2019ApJ...876...58N,2021ApJ...906...72O,2025ApJ...985..158T}), similar to solar flares (e.g., \cite{2000ApJ...540..583S,2019LRSP...16....3T}).
Understanding the occurrence frequency and properties of these superflares is crucial not only for stellar astrophysics but also for evaluating their potential impact on planetary habitability (e.g., \cite{2010AsBio..10..751S,2016NatGe...9..452A,2017MNRAS.465L..34A,2020IJAsB..19..136A,2025kiss.rept.....L}).

Previous statistical studies using Kepler data have shown that the flare frequency increases with the starspot area (e.g., \cite{2013PASJ...65...49S,2013ApJ...771..127N,2019ApJ...876...58N,2021ApJ...906...72O}).
However, most of these studies were ensemble analyses averaging over many stars, or analyses of individual stars over relatively short observational baselines.
Consequently, it remains unclear whether the flare frequency of a single star changes in response to the time evolution of its starspot area (cf.  \cite{2021ApJ...922L..23A}, Supplementary Information in \cite{2022NatAs...6..241N}).
\rev{Moreover, the relationship for a limited number of stars or flares, as well as its dependence on stellar rotation phase (i.e., the visibility of large starspot groups), is not necessarily clear \citep{2023MNRAS.522L..16A,2018ApJ...863..190B,2018ApJ...868....3R,2025ApJ...981..169K}, which may be explained by the presence of polar spots or spot groups distributed over a wide range of longitudes \citep{2014ApJ...797..121H,2023ApJ...948...64I}.}
\rev{While the correlation between flare activity and the 11-year sunspot cycle is well established in the case of the Sun, its verification of ``time-resolved'' relationship in other magnetically active stars remains unexplored.}
The Transiting Exoplanet Survey Satellite (TESS) mission \citep{2015JATIS...1a4003R} provides a unique opportunity to address this issue.
TESS has been performing an all-sky survey since 2018, accumulating photometric data over a baseline of approximately 7 years to date.
This long-term baseline enables us to investigate the time evolution of magnetic activity on individual stars.
In particular, the TESS mission duration now exceeds Kepler’s four-year \rev{prime} mission and therefore offers an advantage for investigating longer-timescale variations, including activity cycles, that could not be explored during the Kepler era.

In this study, we investigate the time-resolved relationship between starspot area and flare frequency on young solar-type stars.
We utilized the long-term optical photometric data from TESS to analyze three representative young solar-type stars: EK Draconis, DS Tucanae A, and V889 Herculis.
Young solar-type stars are of particular interest as they serve as proxies for the ``young Sun,'' offering insights into the magnetic environment of the early solar system.
Also, \rev{Young solar-type stars are also known to be highly flare active, making them suitable targets for studying the temporal evolution of magnetic activity.} 
These three stars are known as bright young solar-type stars and stellar parameters are well known as in Table \ref{tab:targets}, their flare properties and starspot properties are well characterized so far \citep{2005LRSP....2....8B,2005A&A...435..215K,2007LRSP....4....3G,2017MNRAS.465.2076W,2018A&A...620A.162J,2022A&A...661A.148C,2022NatAs...6..241N,2022ApJ...926L...5N,2025ApJ...993...80N,2025arXiv251203830G}.
In this study, by deriving the starspot area and flare occurrence rate for each observational sector, we specifically examine whether the correlation observed in statistical ensemble studies holds for the ``time variation" of individual objects.
Furthermore, for EK Dra, which has the most extensive dataset, we also investigate the dependence of starspot area on the rotation period to explore potential signatures of solar-like differential rotation.
The structure of this paper is as follows: Section \ref{sec:2} describes the target stars and our data analysis methods. Section \ref{sec:3} presents the results of flare detection and the correlation analysis. Section \ref{sec:4} discusses the physical implications of our findings, and we summarize our conclusions in Section \ref{sec:5}.

\section{Data and Analysis}\label{sec:2}


\subsection{TESS Data}\label{sec:tess-data}

TESS performs high-precision photometric monitoring with four wide-field cameras equipped with the TESS filter, which covers the optical band from 6,000 to 10,000 {\AA} \citep{2015JATIS...1a4003R}. Each observing sector spans $\sim$27 days. The data used in this study were retrieved from the Mikulski Archive for Space Telescopes (MAST)\footnote{\url{https://mast.stsci.edu/portal/Mashup/Clients/Mast/Portal.html}}.

TESS observed \rev{V889 Her in four sectors, DS Tuc in five sectors}, and EK Dra in twelve sectors with the 2-minute short-cadence mode. EK Dra was not included in the target list for Sector 50 and was observed only in full-frame images; to maintain uniformity, we excluded this sector from our analysis. Although some sectors provide 20-second cadence data, we did not use them because the higher cadence increases photon noise, which can negatively affect flare detection.
We primarily adopted the Presearch Data Conditioning Simple Aperture Photometry (PDC-SAP) light curves \citep{2015JATIS...1a4003R}. 


\subsection{Flare Detection}

\begin{figure*}
 \begin{center}
  \includegraphics[width=14cm]{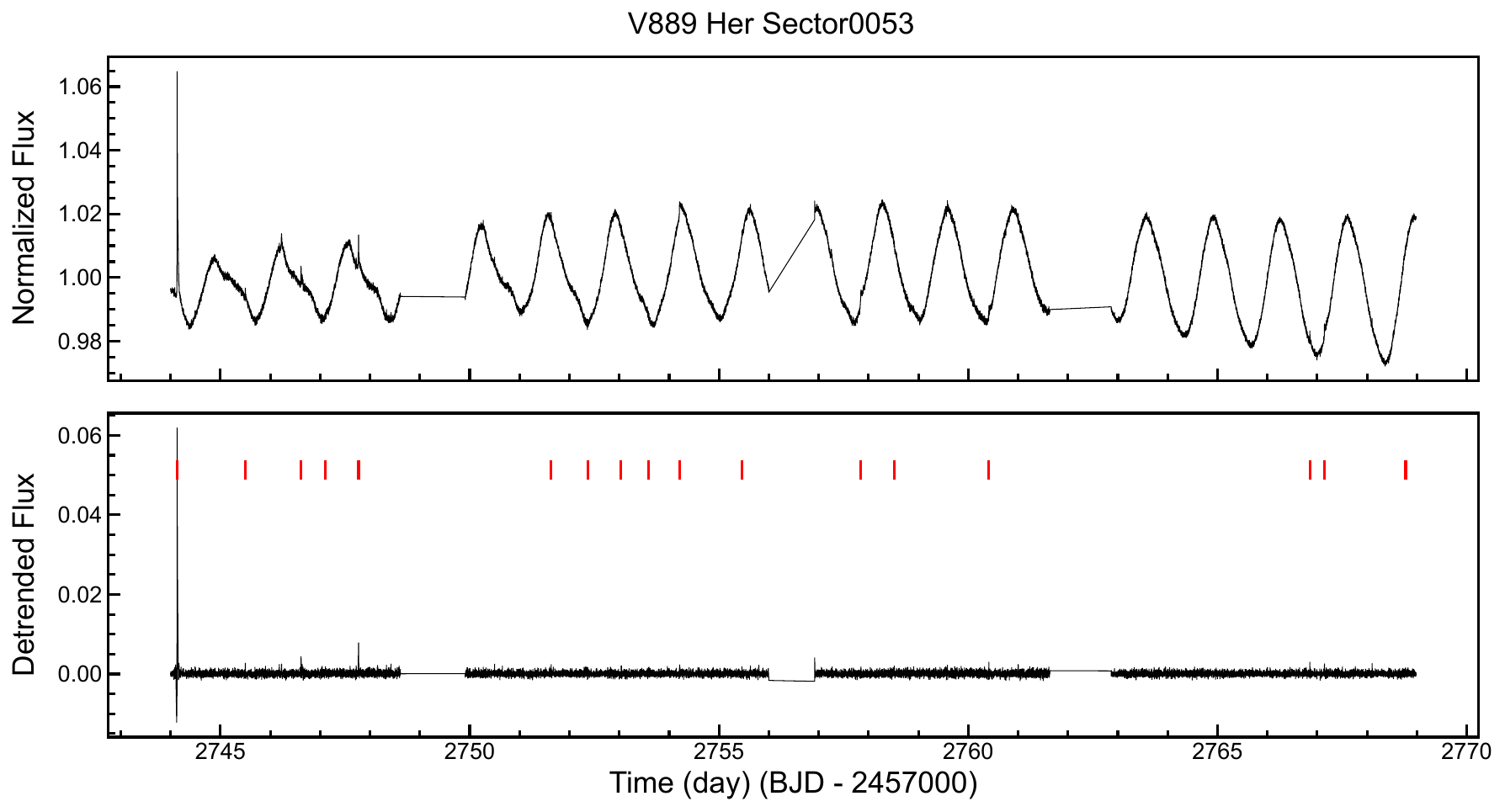} 
 \end{center}
\caption{TESS light curve of V889 Her. (Upper) The light curve normalized by averaged flux \rev{($301519~e^{-}$s$^{-1}$)}. (Lower) The detrened light curve. Red lines marks the timing of detected flares.\\
{Alt text: A two-panel figure for V889 Her. The upper panel shows the TESS light curve normalized by averaged flux and the lower panel shows the detrended light curve. Red markers indicate the timing of detected flares.}
}\label{fig:lc-v889her}
\end{figure*}

\begin{figure*}
 \begin{center}
  \includegraphics[width=14cm]{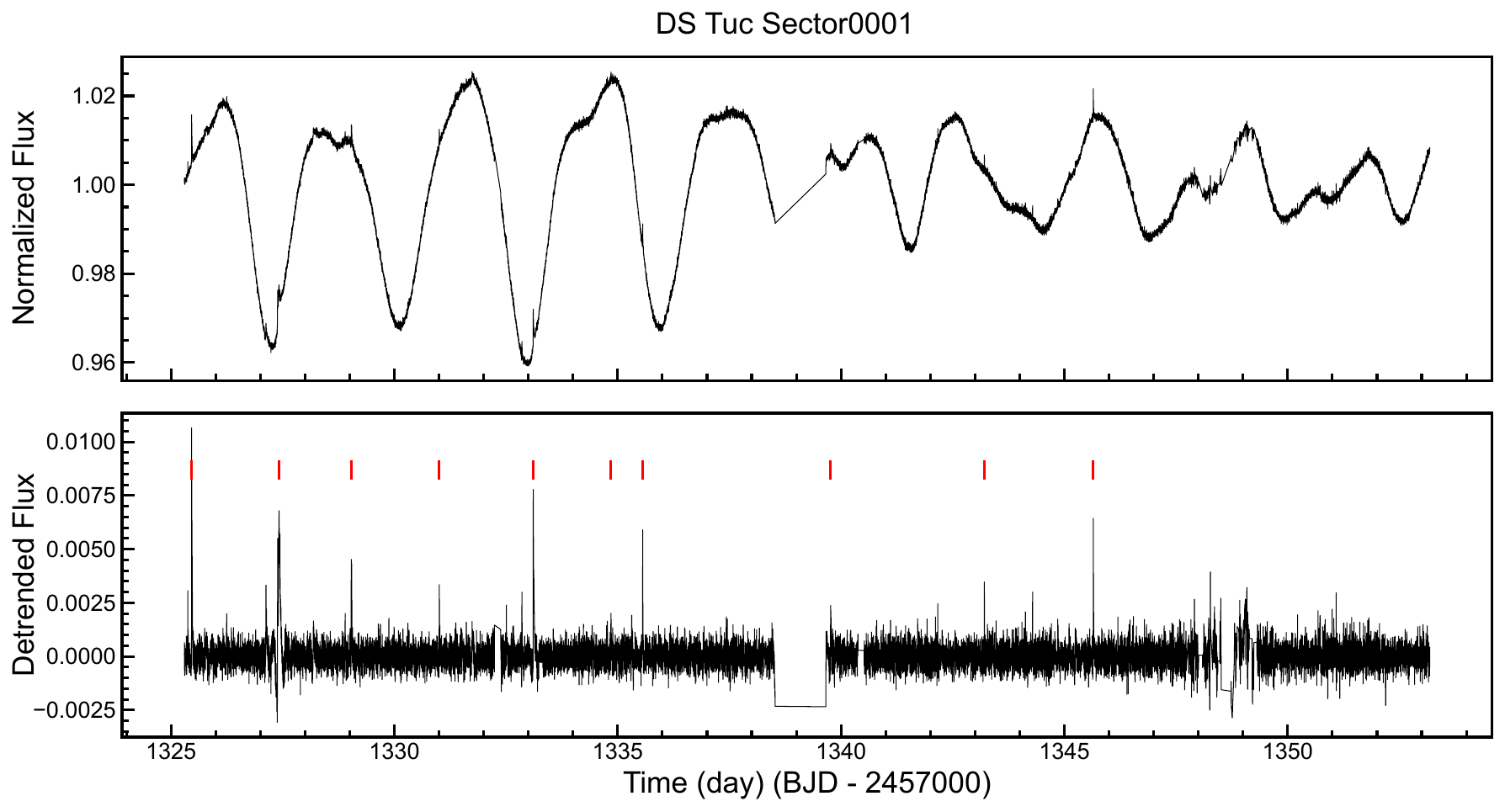} 
 \end{center}
\caption{The same as Figure \ref{fig:lc-v889her} but for DS Tuc. \rev{The averaged flux is ($116892~e^{-}$s$^{-1}$)}\\
{Alt text: A two-panel figure for DS Tuc. The upper panel shows the TESS light curve normalized by averaged flux and the lower panel shows the detrended light curve. Red markers indicate the timing of detected flares.}
}\label{fig:lc-dstuc}
\end{figure*}

\begin{figure*}
 \begin{center}
  \includegraphics[width=14cm]{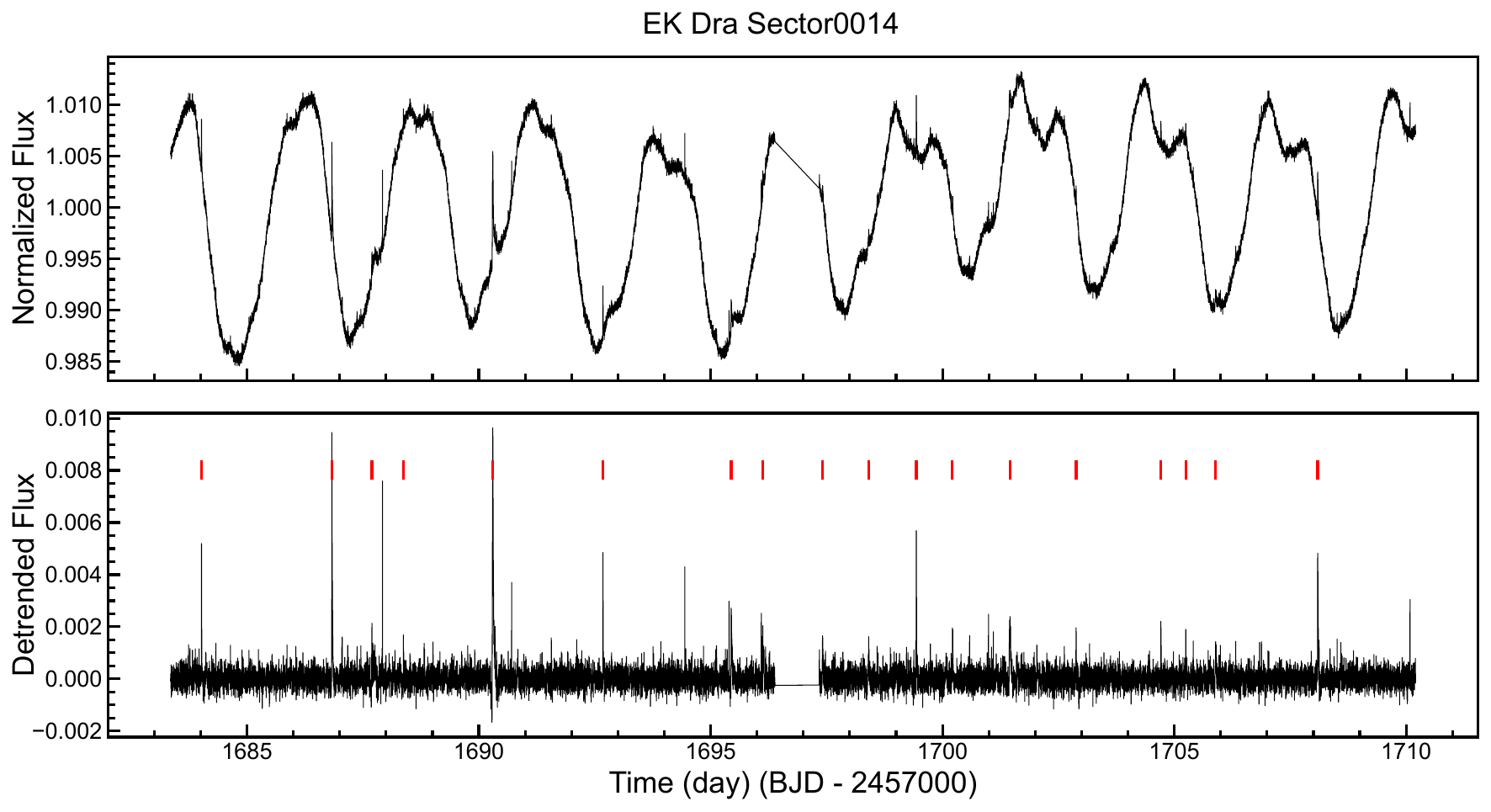} 
 \end{center}
\caption{The same as Figure \ref{fig:lc-v889her} but for EK Dra.\rev{The averaged flux is ($249649~e^{-}$s$^{-1}$)}\\
{Alt text: A two-panel figure for EK Dra. The upper panel shows the TESS light curve normalized by averaged flux and the lower panel shows the detrended light curve. Red markers indicate the timing of detected flares.}
}\label{fig:lc-EKDra}
\end{figure*}

We detected flares from the 2-min cadence TESS light curves using an automated pipeline, basically following the classical method employed in previous papers \citep{2012Natur.485..478M,2022ApJ...926L...5N}.
See Figures \ref{fig:lc-v889her}, \ref{fig:lc-dstuc}, and \ref{fig:lc-EKDra}
for the light curve for each stars.
The analysis code is available here\footnote{\url{https://github.com/yama662607/kyoto-flare-detection}}.
In the following, we briefly summarize the methodology used.

The light curves were processed after normalization by the mean flux. Photometric uncertainties were re-estimated from the local scatter of quiet cadences within a sliding window of $\pm 0.5$~day, and this uncertainty series was used for flare detection.
Flare candidates were identified as sequences of at least two adjacent cadences exceeding $5\sigma$.
\rev{This criterion helps exclude single-cadence brightenings caused by noise, such as cosmic rays.} 
For each candidate, the event window was defined as the contiguous interval where the residual flux exceeds $1\sigma$. Each event was then re-validated using a local linear baseline estimated from pre- and post-event windows, requiring at least two cadences exceeding $3\sigma$ within the refined event.

It should be noted that DS Tuc is a binary system, in which transits can lead to false flare detections and unreliable detrended light curves \citep{2022A&A...661A.148C}. Therefore, the light curve data around the expected transit times were manually removed prior to detrending and flare detection.
In addition, because both binary components are contained within a single pixel in TESS data, flare energies and starspot areas were calculated by taking the binarity into account (Sections \ref{sec:2-3} and \ref{sec:2-4}).

\subsection{Flare Energy}\label{sec:2-3}

We estimate the flare energy in the TESS band using the instrument response function \citep{2015JATIS...1a4003R} and a classical blackbody approximation. A fixed flare blackbody temperature of $T_{\rm flare}=10{,}000$~K is assumed \citep{2013ApJS..209....5S}. This assumption may lead to a few tens of percent error in flare energy \citep{2017ApJ...851...91N}. The time-integrated excess in the normalized light curve is converted to energy using the band-integrated luminosity ratio and the stellar radius.

In the methodology by \citet{2013ApJS..209....5S}, first the flare area is given by
\begin{eqnarray}
A_{\mathrm{flare}}(t) = C_{\mathrm{flare}}(t)\pi R_{\mathrm{star}}^2
\frac{\int R_{\lambda} B_{\lambda}(T_{\mathrm{star}}) {\rm d}\lambda}
{\int R_{\lambda} B_{\lambda}(T_{\mathrm{flare}}) {\rm d}\lambda},
\end{eqnarray}
where $C_{\mathrm{flare}}(t)$ is the relative increase in TESS-band luminosity, $R_{\mathrm{star}}$ is the stellar radius, $R_{\lambda}$ is the TESS response function, and $B_{\lambda}$ is the Planck function. The flare luminosity is then
\begin{equation}
L_{\rm flare}(t) = \sigma_{\rm SB} T_{\rm flare}^4 A_{\rm flare}(t),
\label{eq:L_flare}
\end{equation}
where $\sigma_{\rm SB}$ is the Stefan--Boltzmann constant.
Consequently, the total flare energy is obtained by integrating over the flare duration,
\begin{equation}
E_{\rm flare} = \int L_{\rm flare}(t) {\rm d}t.
\end{equation}
Under this assumption, the determination of flare energies is highly precise. Therefore, we do not assign formal error bars to the energies in this work. However, as noted above, the assumption of a 10,000~K blackbody is not necessarily valid as in M-dwarf flares \citep{2024LRSP...21....1K,2025PASJ...77.1025I} and solar flares \citep{2017ApJ...851...91N}, and systematic uncertainties at the factor-of-a-few level may be present.

\rev{
For DS Tuc, the two stellar components are unresolved in the TESS photometry,
and $C_{\mathrm{flare}}(t)$ is measured relative to the combined
TESS-band flux of the binary system.
We therefore corrected for the binary
dilution by replacing the single-star contribution,
$R_{\mathrm{star}}^2\int R_{\lambda}B_{\lambda}(T_{\mathrm{star}}){\rm d}\lambda$,
with the sum of the TESS-band contributions from DS Tuc A and B. The flare
area for DS Tuc was calculated as
\begin{align}
L_{\mathrm{DS\,Tuc}}
&=
\pi R_{\mathrm{A}}^2
\int R_{\lambda} B_{\lambda}(T_{\mathrm{A}}) {\rm d}\lambda
+
\pi R_{\mathrm{B}}^2
\int R_{\lambda} B_{\lambda}(T_{\mathrm{B}}) {\rm d}\lambda ,
\\
A_{\mathrm{flare}}(t)
&=
C_{\mathrm{flare}}(t)
\frac{
L_{\mathrm{DS\,Tuc}}
}
{
\int R_{\lambda} B_{\lambda}(T_{\mathrm{flare}}) {\rm d}\lambda
} .
\end{align}
}

\rev{Here we note that the assumption of optically thick blackbody emission has not been well validated for solar-type stars, and recent studies mentioned that optically thin emission models may yield energy estimates lower by up to an order of magnitude \citep{2024MNRAS.528.2562S}. However, since this study focuses on the qualitative correlation with starspot area, the main conclusions are unlikely to be affected.}

\subsection{Spot Area}\label{sec:2-4}

Following the method by \citet{2017PASJ...69...41M}, we estimate the starspot area $A_\mathrm{spot}$ from the relative amplitude of rotational modulation, $\Delta F/F$, in the light curve. 
\begin{align}
    A_\mathrm{spot} = \left( \dfrac{\Delta F}{F}\right) A_\mathrm{star}  \frac{T_\mathrm{star}^4}{T_\mathrm{star}^4-T_\mathrm{spot}^4}. \label{eq:A-Tspot}
\end{align}
where $A_\mathrm{star}$ is the stellar projected area.
\rev{For DS Tuc, we estimated the starspot area using the sum of the projected areas of the two stellar components, since both components are contained within a single pixel in the TESS data. Because the TESS photometry cannot distinguish which component hosts the starspots, we assumed the effective temperature of DS Tuc A and adopted the summed projected areas of both components as the effective stellar area.}
We derived the spot temperature $T_{\rm spot}$ from the empirical equation \citep{2005LRSP....2....8B,2017PASJ...69...41M} using the stellar effective temperature $T_\mathrm{star}$:
\begin{align}
T_\mathrm{spot} = - 3.58\times10^{-5} T_\mathrm{star}^2 +0.751 {T_\mathrm{star}} + 808, \label{eq:Tspot-Tphot}
\end{align}
The TESS light curves have very high photometric precision, and the measurement uncertainties are negligible compared to the signal amplitudes. 
However, additional and potentially large uncertainties arise from assumptions such as the unspotted flux level and the presence of multiple spot components (e.g., \cite{2018ApJ...865..142B,2019ApJ...871..187N}). \rev{Therefore, we do not derive error bars to these starspot areas from the light-curve amplitudes. Instead, we estimate their uncertainties by propagating the uncertainties in the stellar effective temperature and radius. Nevertheless, systematic uncertainties would not significantly affect the overall results.}


\rev{Here we note that the equation was revisited by \citet{2021ApJ...907...89H}\footnote{\rev{Here we comment on that the signs of the constant terms in their Equations (5) and (6) in \citet{2021ApJ...907...89H} are incorrect and does not match the observed data. Here we modify the equation as in Equation \ref{eq:Tspot-Tphot-2}}} in the following form:
\begin{align}
T_\mathrm{spot} =  -3.58\times10^{-5} T_\mathrm{star}^2 + 1.0188 {T_\mathrm{star}} - 239.3. \label{eq:Tspot-Tphot-2}
\end{align}
The difference between the inferred spot areas from Equation \ref{eq:Tspot-Tphot-2} and those obtained using Equation \ref{eq:Tspot-Tphot} is just a factor of $\sim$1.2.
This does not change our conclusion.}

\subsection{Effective Observing Time and Flare Frequency}

We computed the effective observing time by multiplying the total number of valid data points used for flare detection by the integration time. Because 2-minute-cadence TESS light curves are used, the effective observing time is defined as the number of data points multiplied by 2 minutes.
The flare frequency is derived by dividing the number of flares, $N_{\rm flare}$, by the effective observing time. The uncertainty in the frequency is estimated assuming Poisson statistics as $\sqrt{N_{\rm flare}+1}$.

\subsection{Rotation Period}

We estimated the stellar rotation period ($P_{\rm rot}$) from the (non-detrended) normalized light curve using the Lomb--Scargle periodogram. The period corresponding to the maximum power is adopted as \rev{the rotation period}. The period uncertainty is estimated by fitting a Gaussian profile to the primary peak of the Lomb--Scargle power spectrum and converting the best-fit width (1$\sigma$ in frequency) into an uncertainty in period.

\section{Result}\label{sec:3}

\subsection{Detected Flares}\label{sssec:3-1}

The top panels of Figures~\ref{fig:lc-v889her}, \ref{fig:lc-dstuc}, and \ref{fig:lc-EKDra} present example light curves of V889~Her, DS~Tuc, and EK~Dra, respectively, with time on the horizontal axis and normalized flux on the vertical axis. The light curves exhibit quasi-periodic variations caused by rotational modulation of starspots \citep{2013ApJ...771..127N,2023ApJ...948...64I}.
The bottom panels show the corresponding detrended light curves, in which these quasi-periodic variations have been removed. In these detrended data, sporadic flux enhancements are interpreted as stellar flares. The red markers indicate the timings of the flares detected by our automated procedures.
The automated procedure successfully identifies the flare events that are also evident by visual inspection in the original light curves.

\begin{figure}
 \begin{center}
  \includegraphics[width=8cm]{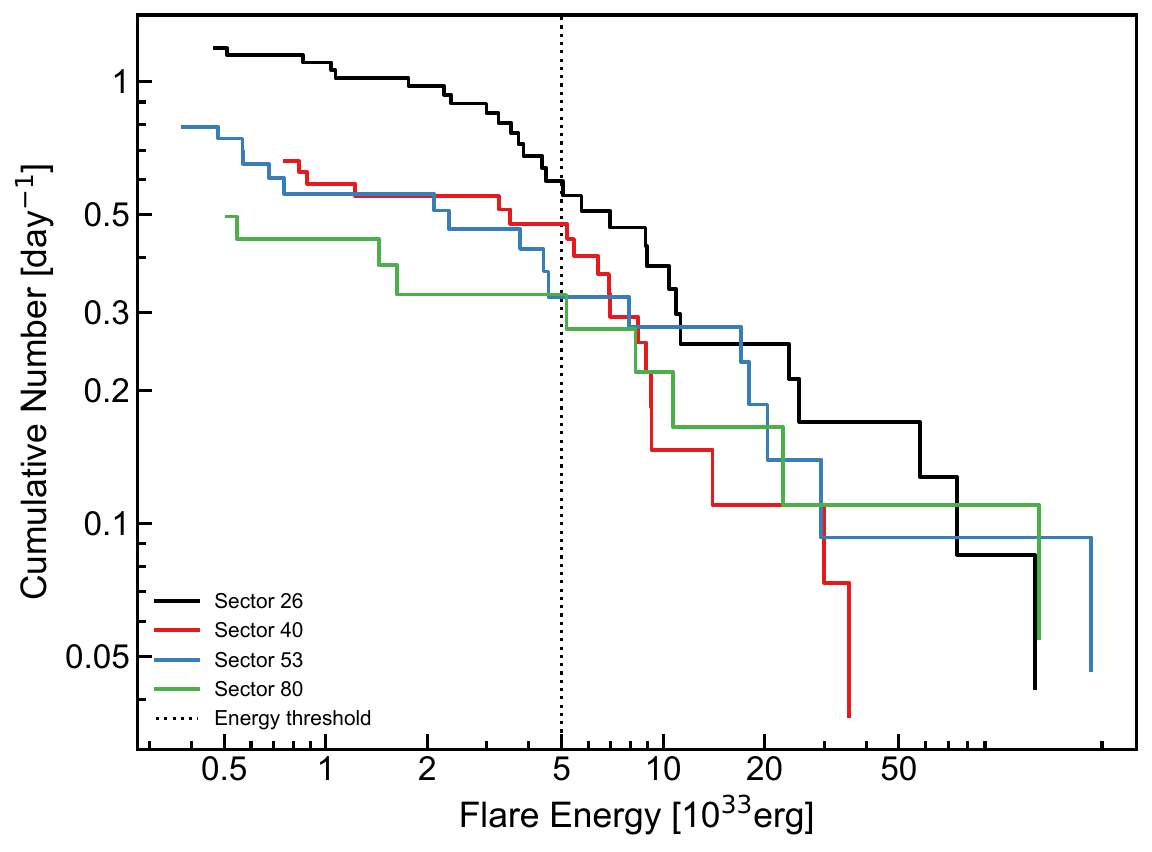} 
 \end{center}
\caption{Cumulative flare energy distribution of V889 Her. Each color represents a different TESS sector. The dashed line indicates the energy threshold of $5\times10^{33}$~erg used to calculate the flare frequency.\\
{Alt text: A cumulative distribution plot of flare energies for V889 Her, with each step-like distribution representing a different TESS sector. A vertical dashed line indicates the energy threshold.} 
}\label{fig:Cumulaive distribution of V889 Her}
\end{figure}

\begin{figure}
 \begin{center}
  \includegraphics[width=8cm]{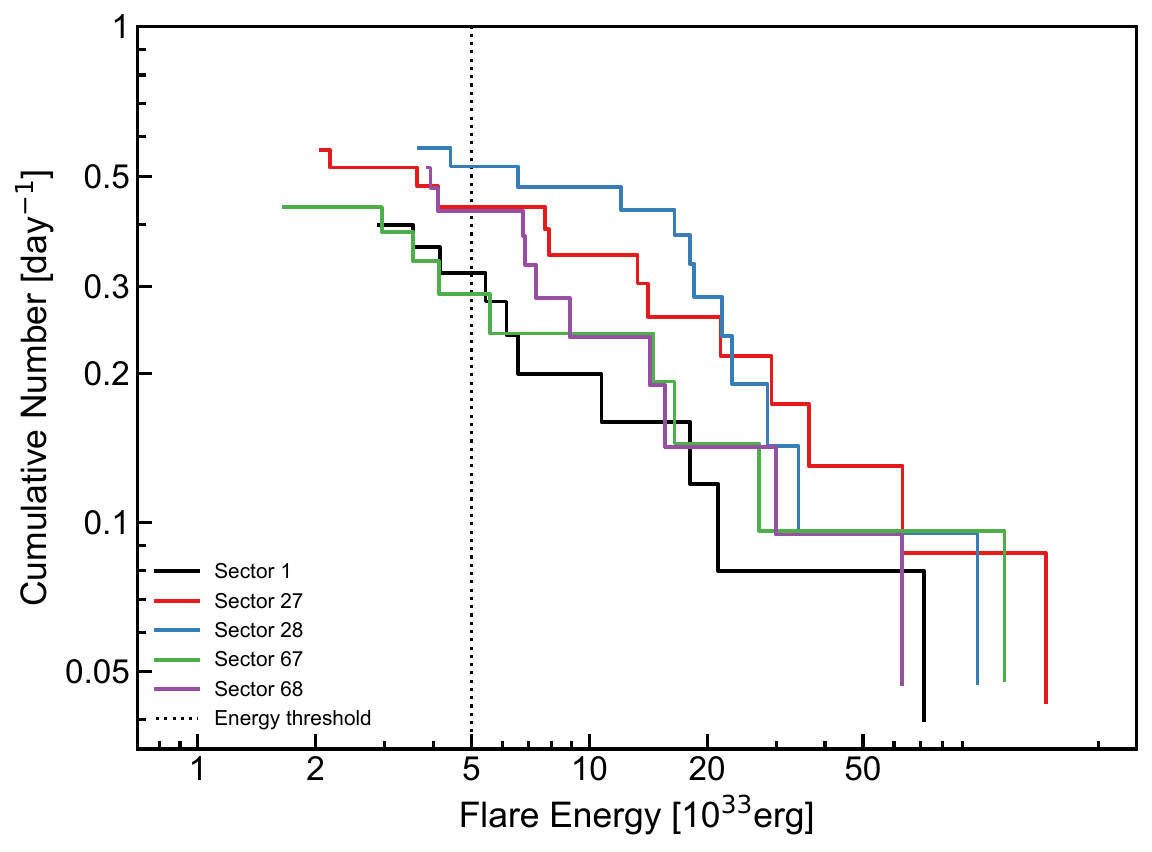} 
 \end{center}
\caption{The same as Figure \ref{fig:Cumulaive distribution of V889 Her} but for DS Tuc.\\
{Alt text: A cumulative distribution plot of flare energies for DS Tuc, with each step-like distribution representing a different TESS sector. A vertical dashed line indicates the energy threshold.}
}\label{fig:Cumulative distribution of DS Tuc}
\end{figure}

\begin{figure}
 \begin{center}
  \includegraphics[width=8cm]{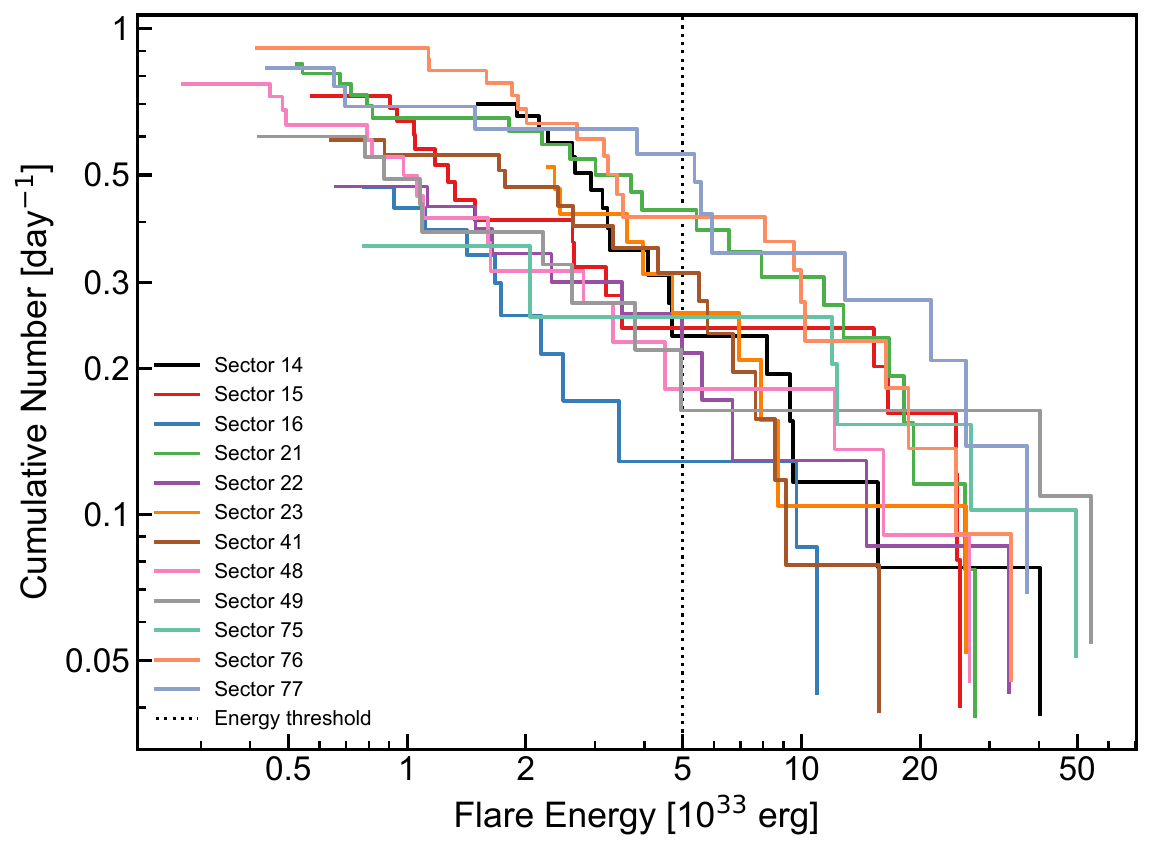} 
 \end{center}
\caption{The same as Figure \ref{fig:Cumulaive distribution of V889 Her} but for EK Dra.\\
{Alt text: A cumulative distribution plot of flare energies for EK Dra, with each step-like distribution representing a different TESS sector. A vertical dashed line indicates the energy threshold.}
}\label{fig:Cumulative distribution of EK Dra}
\end{figure}

Figures~\ref{fig:Cumulaive distribution of V889 Her}, \ref{fig:Cumulative distribution of DS Tuc}, and \ref{fig:Cumulative distribution of EK Dra} show the cumulative flare frequency distributions as a function of bolometric flare energy. The distributions follow power-law forms, while their absolute levels vary from sector to sector (cf. \cite{2022NatAs...6..241N}). The detected number of flares, flare frequency ($>5\times10^{33}$~erg), starspot area, and rotation period are summarized in Table~\ref{tab:flareparam}.

\begin{table}
\caption{Flare parameters.\label{tab:flareparam}}
\centering
\begin{tabular}{ccccc}
\hline
Sect. & N$_{\rm flare}$\footnotemark[*] & Freq$_{\rm flare}$ & $A_{\rm spot}$ & $P_{\rm rot}$ \\
& & [d$^{-1}$] & [$10^{21}$ cm$^{2}$] & [d] \\
\hline
\textsf{V889 Her} \\
26 & \rev{13} & \rev{0.55} & \rev{2.47} & 1.37$\pm$0.03\\
40 & \rev{12} & \rev{0.44} & \rev{1.46} & 1.33$\pm$0.02\\ 
53 & \rev{6} & \rev{0.28} & \rev{2.01} & 1.33$\pm$0.03\\
80 & \rev{5} & \rev{0.28} & \rev{0.94} & 1.41$\pm$0.03\\
\textsf{DS Tuc} \\
1 & 7 & 0.28 & 3.86 & 2.85$\pm$0.14\\ 
27 & 9 & 0.39 & 5.57 & 3.63$\pm$0.20\\ 
28 & 10 & 0.48 & 4.26 & 3.53$\pm$0.19\\ 
67 & 5 & 0.24 & 2.75 & 3.01$\pm$0.12\\ 
68 & 8 & 0.38 & 3.60 & 2.86$\pm$0.13\\ 
\textsf{EK Dra} \\
14 & 5 & 0.19 & 0.86 & 2.64$\pm$0.09\\
15 & 5 & 0.20 &1.20 & 2.67$\pm$0.10\\
16 & 2 & 0.086 & 0.66 & 2.55$\pm$0.09\\ 
21 & 10 & 0.39 & 1.84 & 2.64$\pm$0.09\\ 
22 & 4 & 0.17 & 1.28 & 2.62$\pm$0.09\\ 
23 & 4 & 0.21 & 0.74 & 2.58$\pm$0.09\\ 
41 & 7 & 0.28 & 1.08 & 2.74$\pm$0.10\\ 
48 & 3 & 0.14 & 0.91 & 2.71$\pm$0.10\\ 
49 & 2 & 0.11 & 0.78 & 2.85$\pm$0.11\\ 
75 & \rev{4} & \rev{0.20} & \rev{0.94} & 2.57$\pm$0.09\\ 
76 & \rev{8} & \rev{0.36} & \rev{1.07} & \rev{2.64$\pm$0.11}\\ 
77 & \rev{7} & \rev{0.48} & \rev{1.29} & 2.62$\pm$0.07\\ 
\hline
\end{tabular}
\begin{minipage}{0.5\textwidth}
\footnotesize
\centering
\footnotemark[*]The number of flares with energy $>5\times10^{33}$ erg.
\end{minipage}
\end{table}

\subsection{Relationship between Spot Area and Flare}\label{sssec:3-2}

\begin{figure}
 \begin{center}
  \includegraphics[width=8cm]{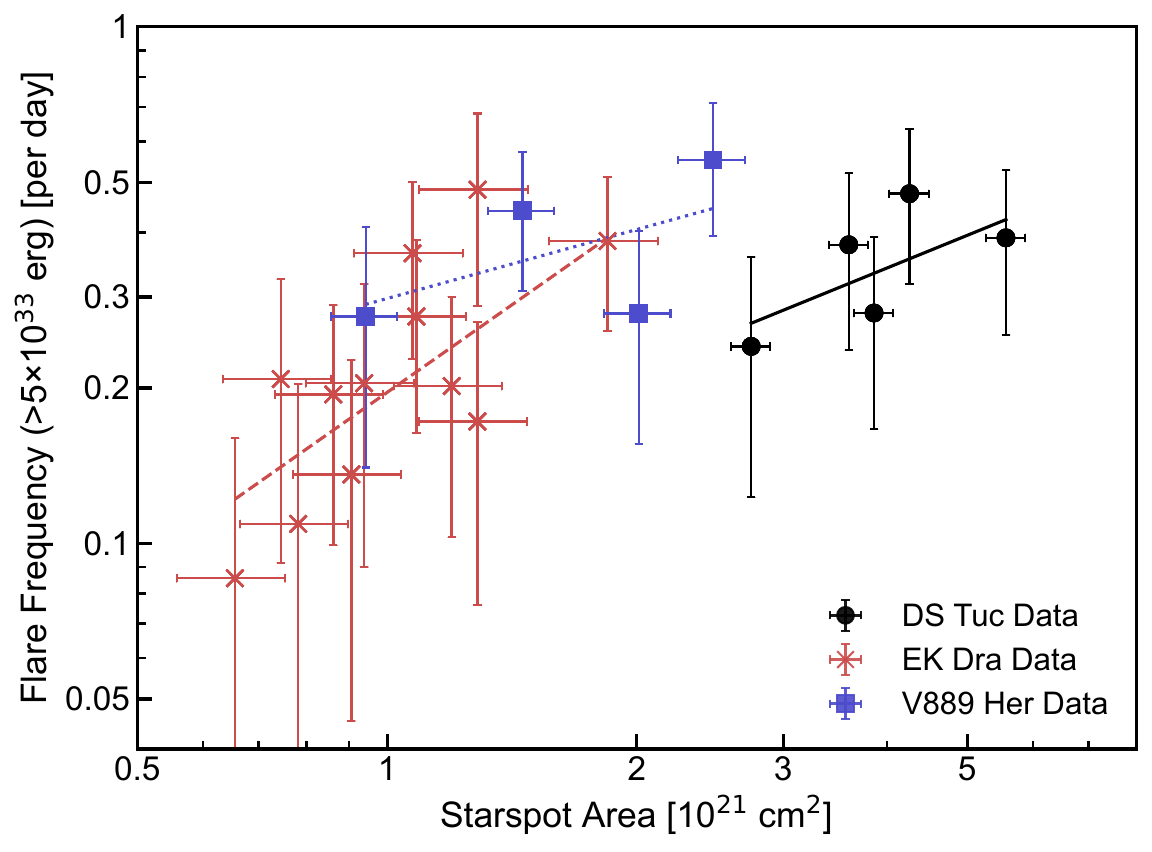} 
 \end{center}
\caption{Relationship between starspot area and flare frequency ($>5\times10^{33}$ erg).  
The blue, black, and red symbols correspond to V889~Her, DS~Tuc, and EK~Dra, respectively. Each point corresponds to a different TESS sector.
Error bars on the flare frequency are Poisson uncertainties, estimated as
$\sigma_{y}=\sqrt{N_{\mathrm{flare}}+1}/T_{\mathrm{obs}}$. 
The data for each star are fitted with a power-law function of the form $y = a x^{b}$ as indicated with a line. 
{Alt text: A scatter plot of flare frequency versus starspot area for DS Tuc, EK Dra, and V889 Her with vertical error bars. DS Tuc points cluster at larger starspot areas on the right, while EK Dra spans smaller to intermediate areas on the left. V889 Her points lie at small to intermediate starspot areas, partially overlapping with the range covered by EK Dra. For each star, larger starspot area is associated with higher flare frequency.}
}\label{fig:spotarea-flarefreq}
\end{figure}

Figure~\ref{fig:spotarea-flarefreq} shows the relationship between the starspot area and the occurrence frequency of flares with energies larger than $5\times10^{33}$~erg, for each star and each TESS sector. 
We found that, for all stars, the flare frequency distributions are affected by incompleteness below $5\times10^{33}$~erg due to limited detection sensitivity (see Figures~\ref{fig:Cumulaive distribution of V889 Her}, \ref{fig:Cumulative distribution of DS Tuc}, and \ref{fig:Cumulative distribution of EK Dra}). We therefore adopt $5\times10^{33}$~erg as a uniform energy threshold.
For all stars, a positive correlation between the starspot area and the flare frequency can be seen. We fitted the data for each star with a power-law function of the form $y = ax^{b}$. 
As a result, we obtained
\rev{$a = 0.30 \pm 0.05$ and $b = 0.44 \pm 0.28$ for V889~Her},
$a = 0.14 \pm 0.06$ and $b = 0.65 \pm 0.33$ for DS~Tuc, and
\rev{$a = 0.20 \pm 0.01$ and $b = 1.13 \pm 0.14$ for EK~Dra}.
\rev{To quantify the statistical significance of these correlations, we also calculated the Pearson correlation coefficients.
The resulting Pearson coefficients and $p$-values are
$r = 0.618$ and $p = 1.1 \times 10^{-2}$ for V889~Her,
$r = 0.598$ and $p = 1.61 \times 10^{-3}$ for DS~Tuc, and
$r = 0.674$ and $p = 2.10 \times 10^{-20}$ for EK~Dra.}
\rev{The resulting Spearman correlation coefficient and $p$-values are
$\rho = 0.800$ and $p = 1.99 \times 10^{-4}$ for V889~Her,
$\rho = 0.800$ and $p = 1.59 \times 10^{-6}$ for DS~Tuc, and
$\rho = 0.650$ and $p = 1.13 \times 10^{-18}$ for EK~Dra.}
We found that the power-law index $b$ is almost unity and common among the stars, while the distribution of flare occurrence frequency at a given starspot area differs from star to star.

\begin{figure}
 \begin{center}
  \includegraphics[width=8cm]{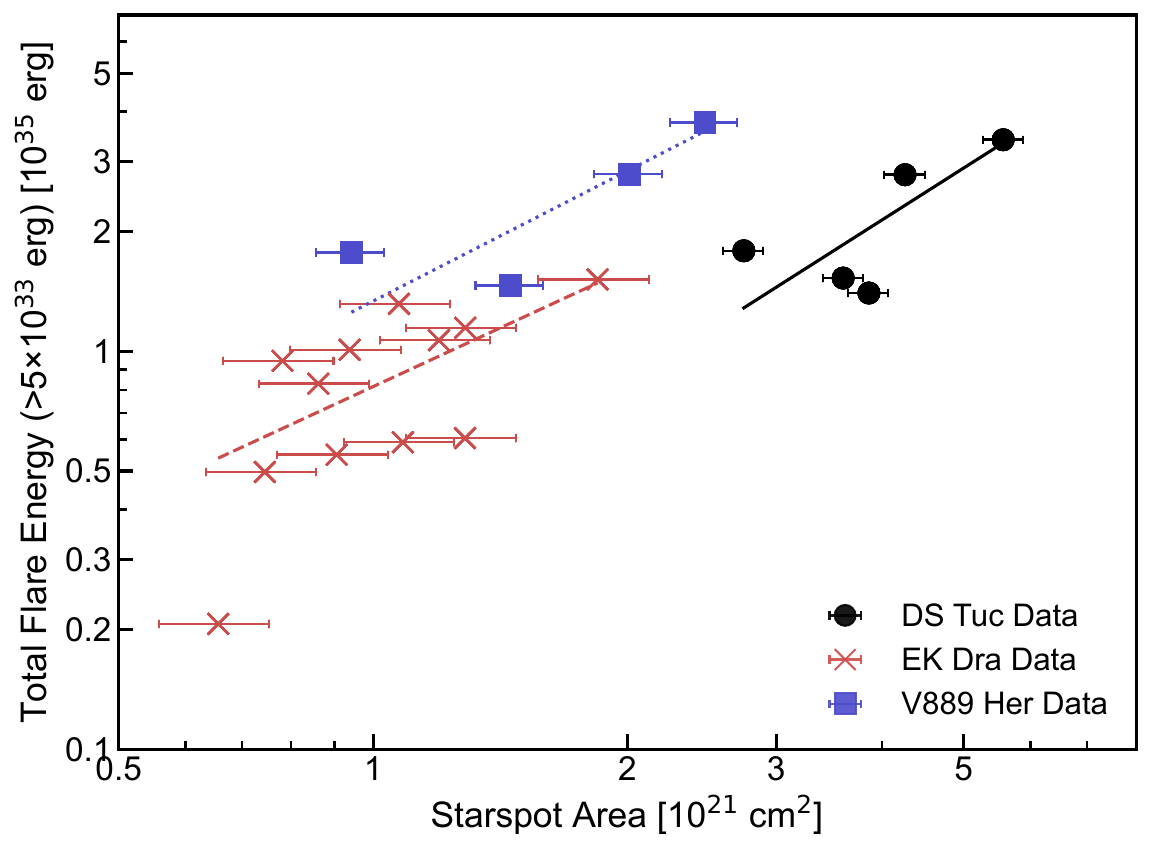} 
 \end{center}
\caption{Relationship between starspot area and total flare energy ($>5\times10^{33}$). The data for each star are fitted with a power-law function of the form $y = a x^{b}$ as indicated with a line.\\
{Alt text: A scatter plot of total flare energy versus starspot area for DS Tuc, EK Dra, and V889 Her. DS Tuc points cluster at larger starspot areas on the right, while EK Dra spans smaller to intermediate areas on the left. V889 Her points lie at small to intermediate starspot areas, showing higher total flare energy than EK Dra at similar starspot areas. For each star, larger starspot area is associated with higher total flare energy.}
}\label{fig:spotarea-totalene}
\end{figure}

Figure~\ref{fig:spotarea-totalene} shows the relationship between the starspot area and the total energy of flares with energies larger than $5\times10^{33}$~erg, plotted in the same manner as in Figure~\ref{fig:spotarea-flarefreq}. 
For all stars, a positive correlation is observed, and the data were fitted with the same power-law function, $y = ax^{b}$. 
The best-fit parameters are
\rev{$a = 1.3 \pm 0.2$ and $b = 1.1 \pm 0.2$ for V889~Her},
$a = 0.33 \pm 0.09$ and $b = 1.4 \pm 0.2$ for DS~Tuc, and
\rev{$a = 0.82 \pm 0.02$ and $b = 0.98 \pm 0.08$ for EK~Dra}.
\rev{The Pearson coefficients and $p$-values are
$r = 0.900$ and $p = 2.06 \times 10^{-6}$ for V889~Her,
$r = 0.813$ and $p = 7.73 \times 10^{-7}$ for DS~Tuc, and
$r = 0.695$ and $p = 4.67 \times 10^{-22}$ for EK~Dra.}
\rev{The Spearman correlation coefficient and $p$-values are
$\rho = 0.800$ and $p = 1.99 \times 10^{-4}$ for V889~Her,
$\rho = 0.600$ and $p = 1.52 \times 10^{-3}$ for DS~Tuc, and
$\rho = 0.699$ and $p = 1.88 \times 10^{-22}$ for EK~Dra.}

While the power-law index $b$ is almost common among the stars, the distribution of the total flare energy as a function of starspot area is also found to differ among the stars.

\begin{figure}
 \begin{center}
  \includegraphics[width=8cm]{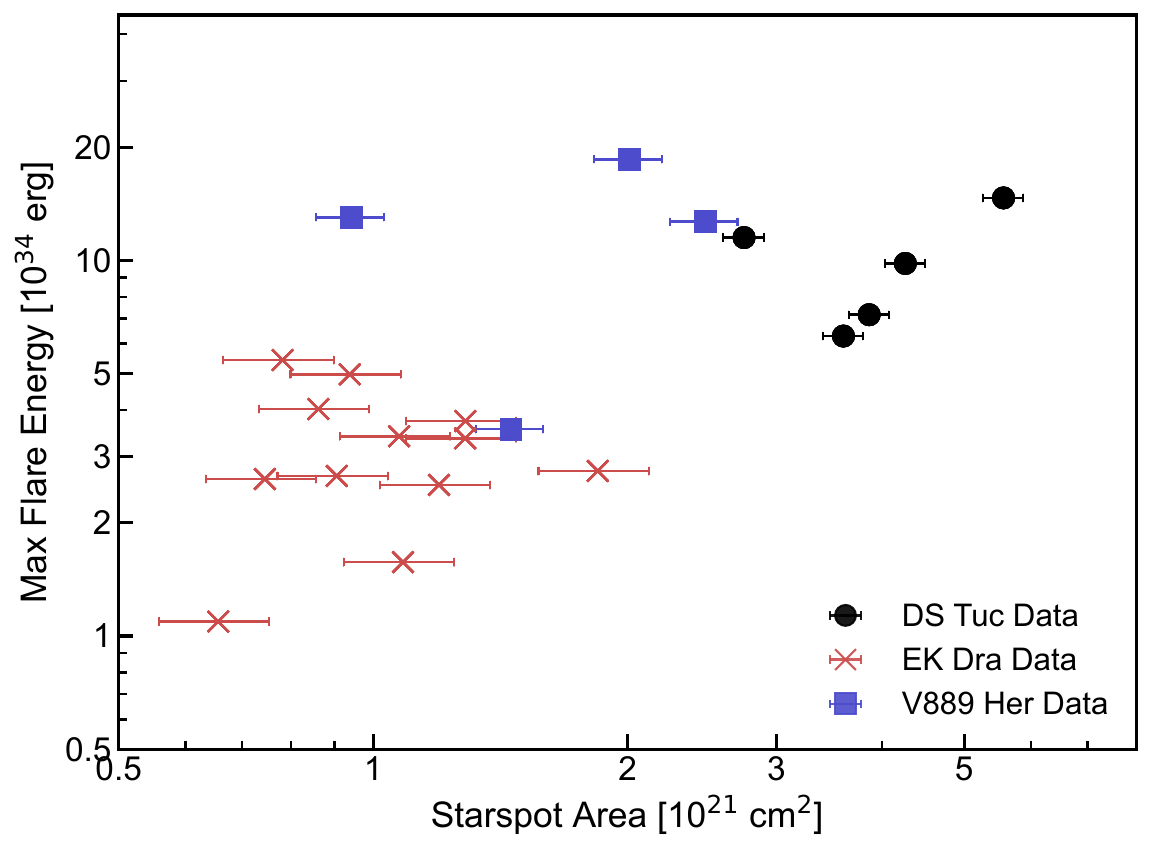} 
 \end{center}
\caption{Relationship between starspot area and maximum flare energy\\
{Alt text: A scatter plot of maximum flare energy versus starspot area for DS Tuc, EK Dra, and V889 Her. DS Tuc points cluster at larger starspot areas on the right, while EK Dra spans smaller to intermediate areas on the left. V889 Her points lie at small to intermediate starspot areas. The maximum flare energy shows no clear trend with starspot area for any of the stars.}
}\label{fig:spotarea-maxene}
\end{figure}

Figure~\ref{fig:spotarea-maxene} shows the relationship between the starspot area and the maximum flare energy observed in each sector, plotted in the same manner as in Figure~\ref{fig:spotarea-flarefreq}. In contrast to Figures~\ref{fig:spotarea-flarefreq} and~\ref{fig:spotarea-totalene}, no clear correlation between the starspot area and the maximum flare energy is found for any of the stars.

\subsection{Starspot Activity as a Function of Rotational Period: Case for EK Dra}

On the present-day Sun, starspots and solar flares occur preferentially at mid-latitudes (cf. \cite{2019LRSP...16....3T}), whereas high-latitude and near-equatorial activity is less common. 
Whether a similar latitudinal preference exists on other active stars remains unclear, although some evidence is obtained for a limited stars (e.g., \cite{2017ApJ...846...99M}). 
Because the Sun exhibits differential rotation, with longer rotation periods at higher latitudes, variations in measured stellar rotation periods may provide indirect information on starspot latitudes if a comparable differential rotation profile applies.

To explore this possibility, we analyzed EK~Dra using long-term photometric data. Rotation periods were derived for each observational sector and compared with the corresponding starspot areas to examine possible latitudinal trends.

\begin{figure}
 \begin{center}
  \includegraphics[width=8cm]{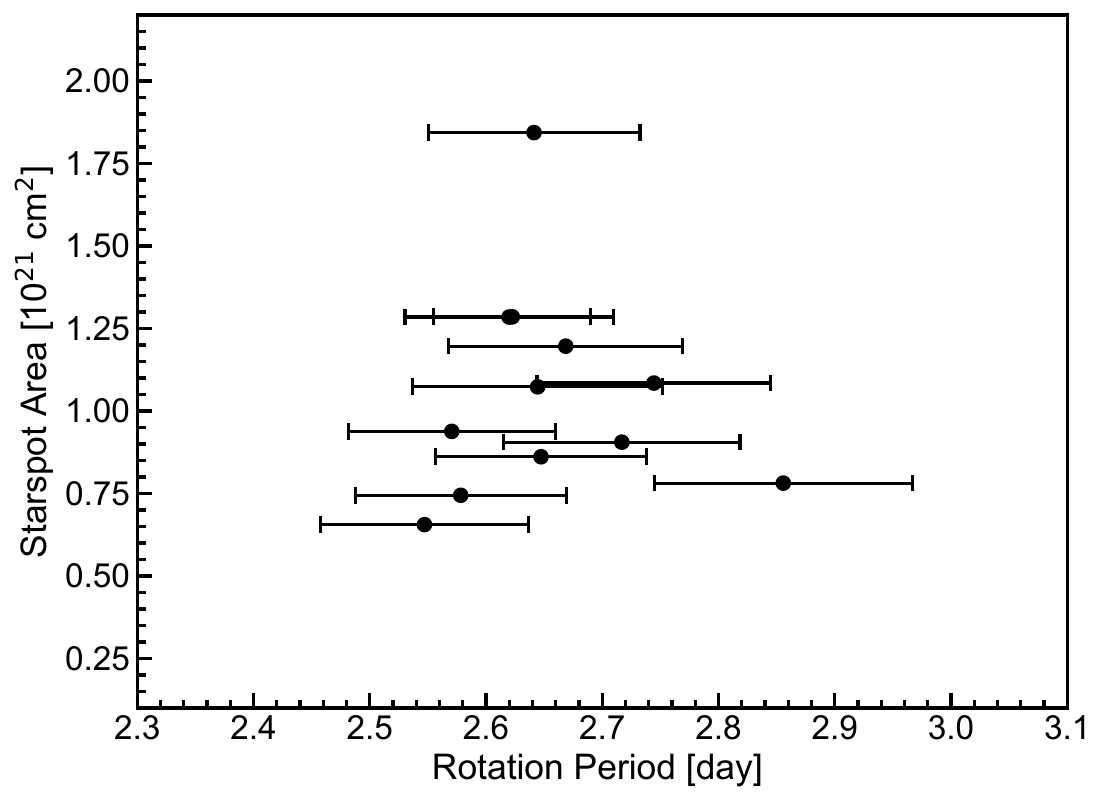} 
 \end{center}
\caption{Relationship between rotational period and starspot area for EK~Dra. Each point corresponds to a different TESS sector.\\
{Alt text: A scatter plot for EK Dra showing starspot area versus rotation period with horizontal error bars. 
Most points lie around intermediate rotation periods (around 2.6–2.7 days), with starspot area showing a broad spread in this range and smaller values at both shorter and longer periods.}
}\label{fig:rotper-spotarea}
\end{figure}

Figure~\ref{fig:rotper-spotarea} shows starspot area as a function of sector-by-sector rotation period. Although the period uncertainties are large, starspot areas appear to be enhanced at intermediate rotation periods. This tentative trend suggests that, as on the Sun, larger starspots on EK~Dra may preferentially emerge at mid-latitudes.
\rev{The estimated differential rotation $\Delta\Omega \approx 2\pi \left(1/P_{\min} - 1/P_{\max}\right)$ gives $\sim0.25~{\rm rad\,day^{-1}}$, which is generally consistent with Kepler-based studies (e.g., \cite{2016MNRAS.461..497B}), which derived surface differential rotation from photometric frequency splittings.}

\rev{Here we note that the interpretation of differential rotation in EK~Dra may require caution. The variation in $P_{\rm rot}$ could result from starspot evolution, morphological changes, or shifts in dominant active regions rather than latitude changes (e.g., \cite{2014A&A...564A..50L,2016MNRAS.461..497B}). }


\section{Discussion}\label{sec:4}

Figures~\ref{fig:spotarea-flarefreq} and \ref{fig:spotarea-totalene} show that both the flare occurrence rate and the total flare energy exhibit positive correlations with starspot area. This behavior closely resemble the well-established relationship between solar flares and sunspots \citep{2000ApJ...540..583S,2019LRSP...16....3T}, and indicates that, on young solar-type stars as well, superflares are powered by the release of magnetic energy stored in large starspot groups.
These correlations further imply that temporal variations in starspot area (i.e., the amount of stored magnetic energy) can lead to substantial changes in the flare occurrence rate (i.e., the magnetic energy release rate). We note, however, that no clear signatures of activity-cycle modulation are detected in our sample. \rev{Although an $\sim$8-yr activity cycle has previously been reported for EK Dra \citep{2025arXiv251203830G}, the sparse sampling and limited TESS baseline make it difficult to give a conclusive evidence of activity cycles.}

Figures~\ref{fig:spotarea-flarefreq} and \ref{fig:spotarea-totalene} further reveal that the distributions differ among the three stars. In particular, DS~Tuc tends to exhibit either a larger inferred starspot area or a lower estimated flare frequency compared to EK~Dra and V889~Her (with V889~Her possibly showing a slightly lower flare frequency than EK~Dra).
We interpret these differences primarily as observational and methodological effects. First, the TESS light curve of DS~Tuc contains flux contributions from both DS~Tuc~A and DS~Tuc~B. Consequently, starspot areas derived under the assumption of a single star are likely overestimated. Second, the combined flux from the two components increases the mean system brightness, reducing the relative amplitude of individual flare brightenings compared to those in single-star systems such as EK~Dra and V889~Her. This dilution makes low-energy flares more difficult to detect and can lead to an underestimation of the flare occurrence rate.
In addition, DS~Tuc is more distant than the other two targets (Table~1), and therefore has a lower apparent brightness. As a result, photon-counting statistics and instrumental noise become more important, further decreasing the detection efficiency for low-energy flares.
\rev{Another possible factor is that the energy threshold ($5 \times 10^{33}$) may lead to an underestimation of the flare number, particularly in Sector 28 (see the blue line in Figure \ref{fig:Cumulative distribution of DS Tuc}).}
Taken together, these effects provide a natural explanation for the systematically lower apparent flare frequency and larger inferred spot area of DS~Tuc relative to the other stars. \rev{Specifically, the binarity of DS~Tuc may cause the starspot area to be overestimated, whereas its larger distance from the Earth may reduce the flare-detection completeness and thus lead to an underestimated flare occurrence rate. Therefore, these effects are likely to introduce systematic offsets in the inferred quantities, rather than implying a physical modification of the underlying power-law relation.}

\rev{
We also derived a power-law relation between the starspot area and the activity index. 
Assuming that the spot area scales as $A_{\rm spot} \sim L^{2}$, the integrated activity indicator, such as the flare frequency and total flare energy, is expected to reflect the total free magnetic energy stored in the active region. 
This can be written as
\begin{equation}
f E_{\mathrm{mag}} \sim \frac{1}{8\pi} B^{2} L^{3} \propto A_{\rm spot}^{3/2},
\label{eq:mag}
\end{equation}
where $f$ is the filling factor, $B$ is the magnetic field strength, and $L$ is the characteristic length scale.
The observationally derived power-law indices can therefore be compared with the theoretical index of 1.5. 
The observed indices of $0.44\text{--}1.4$ are somewhat smaller, but broadly consistent with this expectation. 
However, given the limited dynamic range and the substantial scatter in the data, caution is required when interpreting these results.
}

Finally, we find no clear correlation between starspot area and the maximum flare energy. One plausible explanation is that photometric light curves alone cannot distinguish whether the observed rotational modulation is produced by a single exceptionally large spot group or by an ensemble of multiple moderately sized spots. As a result, the inferred spot area does not necessarily reflect the size of the largest magnetic structure capable of powering the most energetic flares.
Moreover, as shown in Figures~\ref{fig:Cumulaive distribution of V889 Her}--\ref{fig:Cumulative distribution of EK Dra}, the highest-energy flares are intrinsically rare, and the number of such events in our sample is limited. Consequently, statistical uncertainties are large and stochastic effects due to small-number statistics cannot be ruled out. Together, these factors likely contribute to the absence of an apparent correlation between starspot area and maximum flare energy.

\begin{figure}
\begin{center}
\includegraphics[width=8cm]{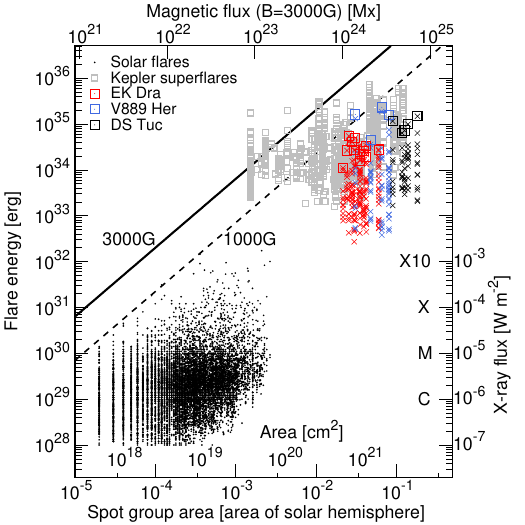}
\end{center}
\caption{\rev{Relationship between starspot area and flare energy, overplotted with superflares on Kepler solar-type stars (gray; $T_{\rm eff}=$5600--6000 K) and solar flares (black) from \citet{2021ApJ...906...72O}.
The relation of $E_{\rm{flare}} = \frac{1}{8\pi} f B^2 A_{\rm{spot}}^{3/2}$ (e.g., \cite{2013PASJ...65...49S}, see also text)
is shown for magnetic field strengths of 1000 G (dashed line) and 3000 G (solid line). Individual flares identified in this study are shown as crosses for EK Dra (red), V889 Her (blue), and DS Tuc A (black), while the maximum flare energy in each sector is indicated by squares. Note that the apparent excess above the 1000 G relation for Kepler solar-type stars may be attributed to low inclination angles inferred from spectroscopic studies \citep{2015PASJ...67...33N,2019ApJ...876...58N}, which can lead to an underestimation of the starspot area.} \\
{Alt text: Log--log plot of flare energy versus starspot area comparing solar flares, Kepler superflares, and flares from EK Dra, V889 Her, and DS Tuc A, with reference scaling relations for different magnetic field strengths.}}
\label{fig:spotarea-maxene:kepler}
\end{figure}

\rev{
Theoretically, a correlation is also expected between starspot area and the maximum flare energy, as the upper limit of the released energy is constrained by the free magnetic energy described described in Figure \ref{eq:mag} (e.g., \cite{2013PASJ...65...49S}). 
Previous studies have shown that the upper envelope of flare energies, for both solar flares and superflares on solar-type stars, can be statistically explained by Equation \ref{eq:mag} (e.g., \cite{2013PASJ...65...49S,2013ApJ...771..127N,2019ApJ...876...58N,2021ApJ...906...72O}). 
Indeed, as shown in Figure~\ref{fig:spotarea-maxene:kepler}, our data are broadly consistent with the expected scaling for magnetic field strengths comparable to those on the Sun.
Therefore, the absence of a clear correlation between starspot area and maximum flare energy in our sample (Figure~\ref{fig:spotarea-maxene}) does not necessarily contradict the theoretical expectation, but is more likely attributable to limited statistics and the narrow dynamic range of the present data.
}

\section{Summary and Conclusion}\label{sec:5}

We investigated the time-resolved relationship between starspot area and flare activity on individual young solar-type stars using $\sim$7 years of TESS photometry. Focusing on V889~Her, DS~Tuc, and EK~Dra, we automatically detected flares and derived the flare frequency, starspot area, and rotation period for each $\sim$27-day TESS sector.

We find that both the starspot area and the flare frequency vary substantially from sector to sector, although no clear activity-cycle-like modulation is detected. For all three stars, the flare occurrence rate above $5\times10^{33}$~erg exhibits a positive correlation with starspot area, and the relation is well described by a power law with an index close to unity. This result provides direct, time-resolved evidence that stellar flare activity on young solar-type stars is causally linked to the amount of magnetic energy stored in starspots, closely analogous to the solar case. 
We find no clear correlation between starspot area and the maximum flare energy \rev{among stars, but the absolute values of maximum flare energy can be well explained by a common scaling relation with solar flares and sunspots. The lack of strong correlation} is plausibly explained by the inability of photometric light curves to isolate the largest individual spot groups, together with the intrinsically low occurrence rate of the most energetic flares and resulting small-number statistics.
For EK~Dra, we further examined the relationship between starspot area and sector-by-sector rotation period and found a tentative enhancement of spot area at intermediate rotation periods. Assuming differential rotation, this may indicate that large starspots preferentially form at mid-latitudes, similar to the Sun, although the current uncertainties prevent a definitive conclusion.

Overall, our results indicate that the fundamental connection between starspots and flares established for the Sun extends to the extreme magnetic activity of young solar-type stars. Future progress will require simultaneous spectroscopic mapping of starspots (e.g., Doppler imaging, cf. \cite{2025arXiv251203830G,2025arXiv251112190L}) and continued long-term photometric monitoring, as well as expanding the sample to a larger population of Kepler and TESS targets, in order to place stronger constraints on the spatial distribution of starspots and the long-term evolution of the spot–flare connection.

\begin{ack}
\rev{We thank the referee for their careful review, which has helped improve the quality of our manuscript. We thank Prof. Shibata Kazunari and Dr. Hiroyuki Maehara for their fruitful comments.}
This work was supported by JSPS (Japan Society for the Promotion of Science) KAKENHI Grant Numbers 21J00316, 24H00248, 24K00680, and 25K01041 (K.N.). 
This work was supported by the Operation Management Laboratory (OML) of the National Institutes of Natural Sciences (NINS), Japan (K.N.). 
This paper includes data collected with the TESS mission, obtained from the MAST data archive at the Space Telescope Science Institute (STScI). Funding for the TESS mission is provided by the NASA Explorer Program. STScI is operated by the Association of Universities for Research in Astronomy, Inc., under NASA contract NAS 5-26555. 
The authors acknowledge ideas from the participants in the workshop ``Blazing Paths to Observing Stellar and Exoplanet Particle Environments" organized by the W.M. Keck Institute for Space Studies.
The authors also would like to acknowledge the the relevant discussions in the International Space Science Institute (ISSI)
Workshop ``Stellar Magnetism and its Impact on (Exo)Planets (\url{https://workshops.issibern.ch/stellar-magnetism/})".
\end{ack}



\end{document}